\documentclass{article}
\usepackage{spconf}
\usepackage{graphicx}
\usepackage{amsmath}
\usepackage{cite}
\usepackage{booktabs}
\usepackage[draft=false,colorlinks=true]{hyperref}
\usepackage{caption}
\usepackage{subcaption}
\usepackage{tablefootnote}

\title{On the role of Lip Articulation in  Visual Speech Perception}
\name{%
\begin{tabular}{c}
Zakaria Aldeneh\sthanks{Authors contributed equally.},
Masha Fedzechkina$^*$,
Skyler Seto,
Katherine Metcalf,\\
Miguel Sarabia,
Nicholas Apostoloff,
Barry-John Theobald
\end{tabular}}
\address{Apple}
\begin{document}
%
\maketitle
\begin{abstract}
Generating realistic lip motion from audio to simulate speech production is critical for driving natural character animation. Previous research has shown that traditional metrics used to optimize and assess models for generating lip motion from speech are not a good indicator of subjective opinion of animation quality. Devising metrics that align with subjective opinion first requires understanding what impacts human perception of quality. In this work, we focus on the degree of articulation and run a series of experiments to study how articulation strength impacts human perception of lip motion accompanying speech. Specifically, we study how increasing under-articulated (dampened) and over-articulated (exaggerated) lip motion affects human perception of quality. We examine the impact of articulation strength on human perception when considering only lip motion, where viewers are presented with talking faces represented by landmarks, and in the context of embodied characters, where viewers are presented with photo-realistic videos. Our results show that viewers prefer over-articulated lip motion consistently more than under-articulated lip motion and that this preference generalizes across different speakers and embodiments.
\end{abstract}
%

\section{Introduction}
Animating faces from speech has applications in interactive systems, entertainment, and accessibility. Most recent approaches for generating visual speech use neural networks, especially recurrent neural networks (RNNs)~\cite{eskimez2018generating,greenwood2018joint,wiles2018x2face,zakharov2019few,jamaludin2019you,wang2021one}, by training the model to replicate ground-truth reference visual speech from acoustic speech features.  Objective metrics typically used to optimize these models, such as the mean squared error (MSE), capture errors globally and can fail to capture \emph{perceptually} significant errors~\cite{theobald2012relating,hussen2020modality,websdale2021speaker}.  Thus, assessing trained models usually involves some form of \textit{subjective} assessment in the form of pairwise preference testing or mean opinion score (MOS) aggregation~\cite{taylor2017deep,suwajanakorn2017synthesizing,zakharov2019few,hussen2019speaker,zhou2020makelttalk,vougioukas2020realistic,ji2021audio,hussen2021audiovisual,websdale2021speaker}, which are time-consuming and expensive.


Previous work has highlighted that commonly used objective measures for assessing generated visual speech quality are not indicative of subjective opinion of quality~\cite{theobald2012relating} as sequences with little difference in MSE can vary considerably in terms of subjective opinion of quality~\cite{hussen2020modality}.  At issue is the \emph{type} of error and \emph{where} in the sequence errors occur are not taken into account. For example, missing a lip closure during a bilabial plosive (/b/) is more significant perceptually than parting the lips slightly more than is necessary for a velar plosive (/k/).  Consequently, several approaches have been proposed to address the limitations of traditional objective metrics. For example, Zhou et al.~\cite{zhou2020makelttalk} proposed a collection of metrics computed from various distances in lip/jaw position and velocity. Alternatively, Prajwal et al.\ used a pre-trained SyncNet model to quantify the quality of visual speech~\cite{prajwal2020lip}. Chen et al.~\cite{chen2020comprises} used the distance between embeddings from a pre-trained lip reading model extracted from real sequences and generated equivalents as a measure of quality. Although previously proposed approaches can mitigate the limitations of traditional metrics, none of the approaches bridge the disconnect between human evaluations and objective measures.

Devising objective metrics that align with perceived quality of visual speech necessarily involves understanding how sources of error influence the perception of quality. In this work, we focus on the degree of articulation and ask whether the perceived quality of visual speech is influenced more by under-articulated (dampened) lip motion or by over-articulated (exaggerated) lip motion. We first study this effect from the perspective of pure motion by rendering the visual speech as facial landmarks, akin to point-light displays used to understand speech production. Then, we examine the impact of the degree of articulation in the context of a photo-realistic talking face created by a state-of-the-art approach~\cite{zhou2020makelttalk}. We find that over-articulation impacts the perceived quality much less than under-articulation does. We conclude with a discussion about using human perception of visual speech quality to improve model development and evaluation.


\begin{figure*}[ht]
  \centering
  \begin{tabular}{cccccc}
  \includegraphics[scale=0.175, clip, trim=4cm 2cm 4.5cm 10cm]{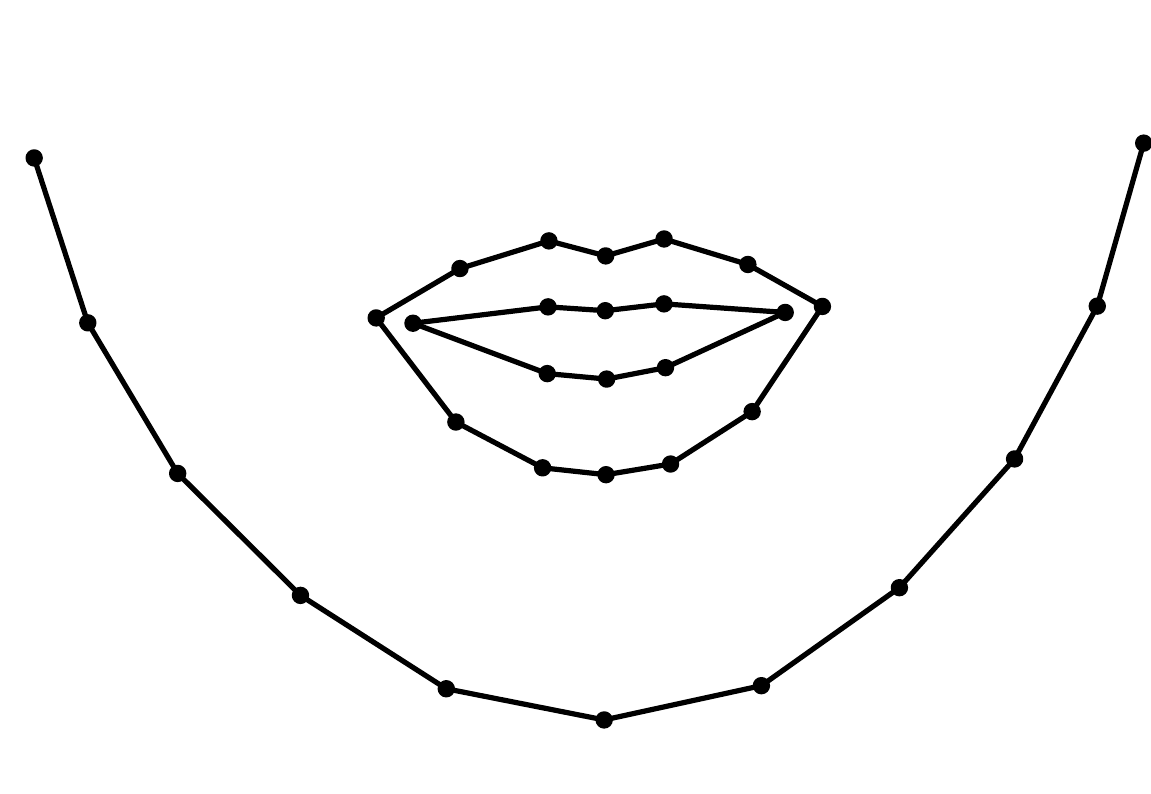} & 
  \includegraphics[scale=0.175, clip, trim=4cm 2cm 4.5cm 10cm]{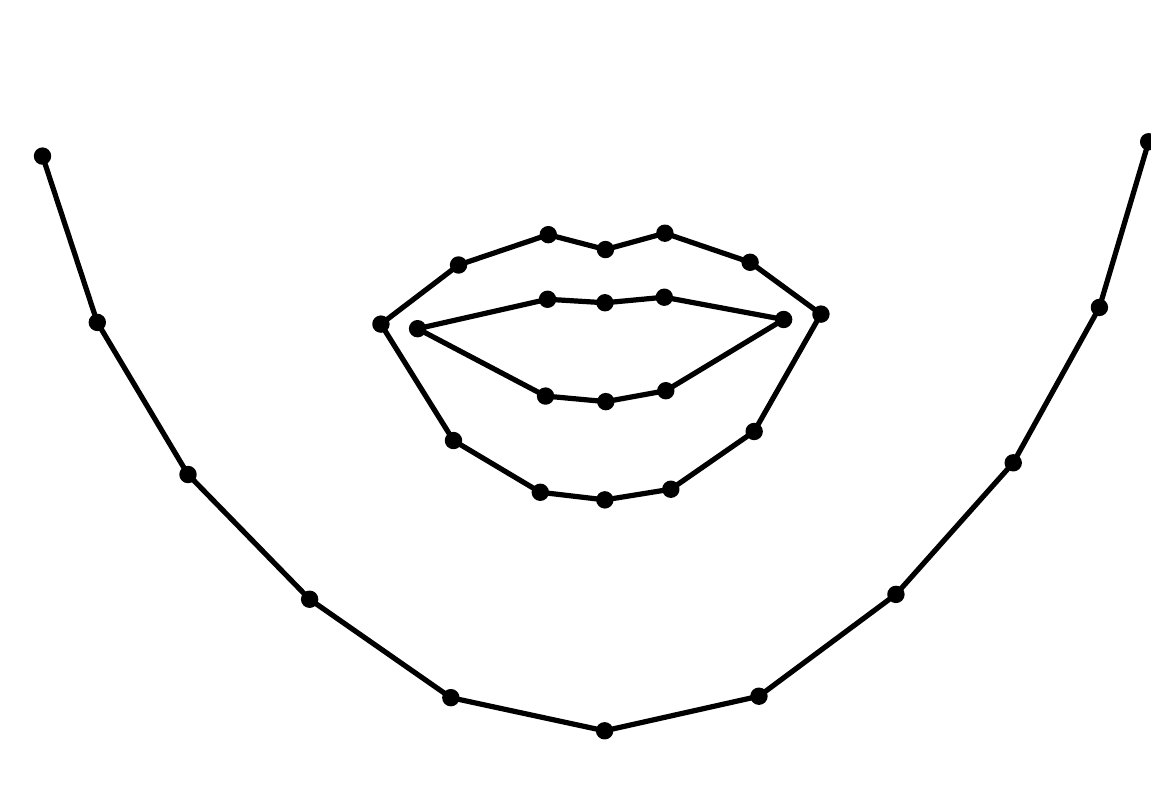} &
  \includegraphics[scale=0.175, clip, trim=4cm 2cm 4.5cm 10cm]{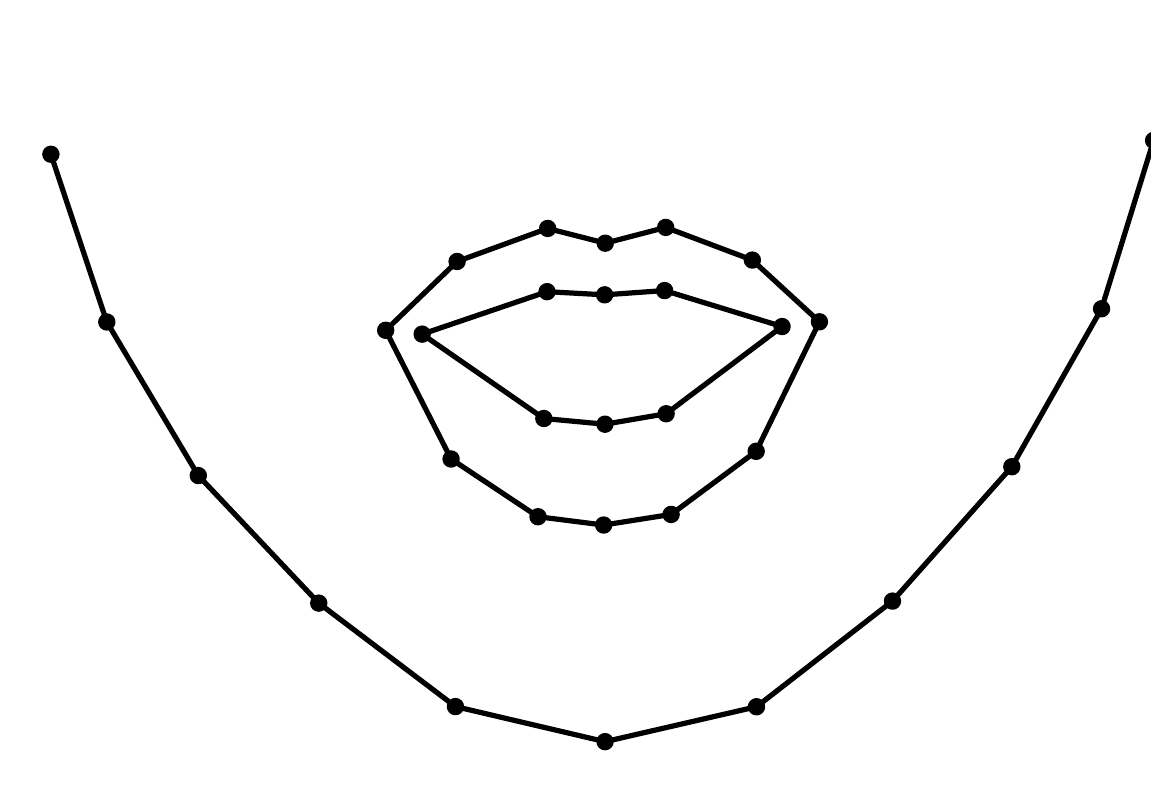} &
  \includegraphics[scale=0.175, clip, trim=4cm 2cm 4.5cm 10cm]{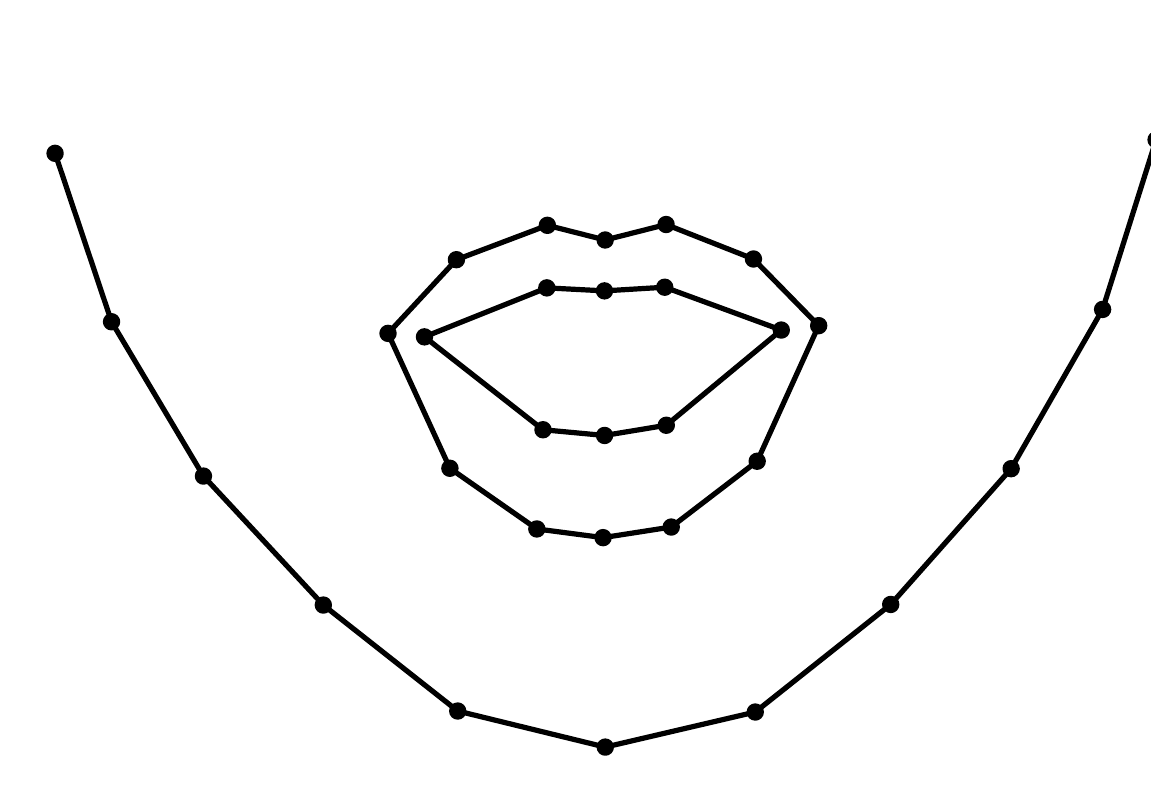} &
  \includegraphics[scale=0.175, clip, trim=4cm 2cm 4.5cm 10cm]{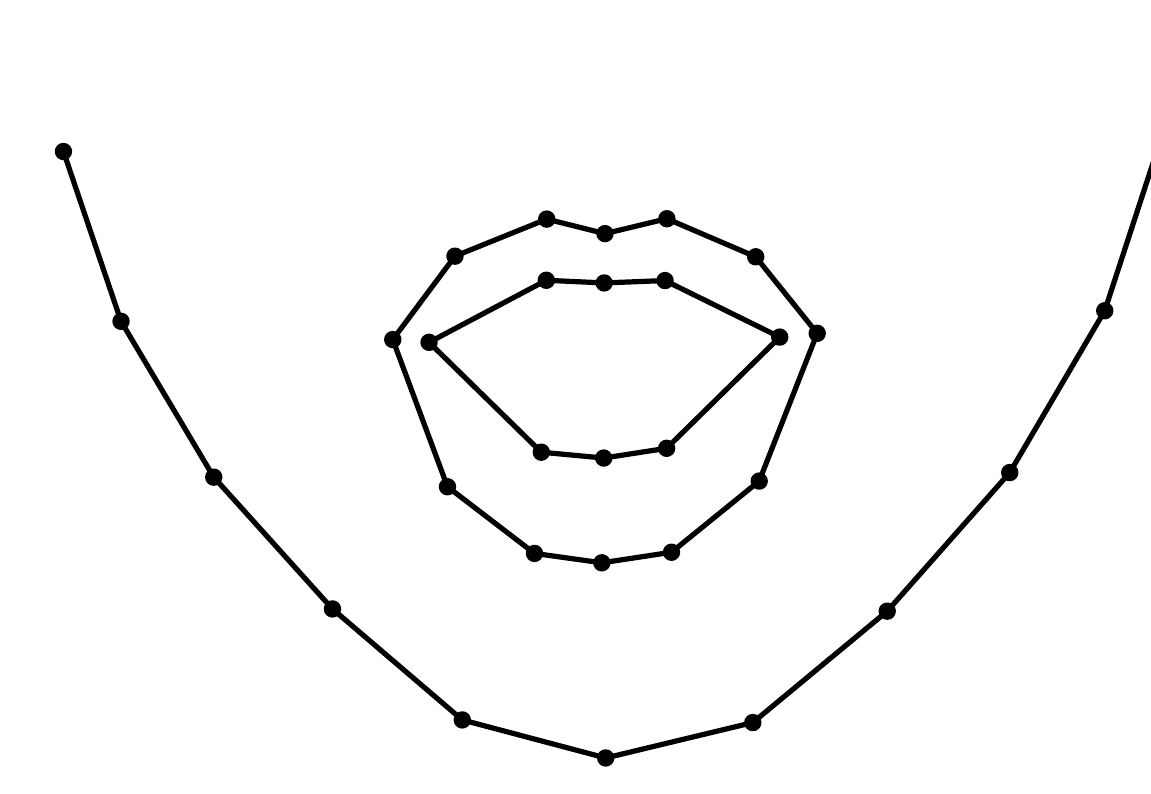} &
  \includegraphics[scale=0.175, clip, trim=4cm 2cm 4.5cm 10cm]{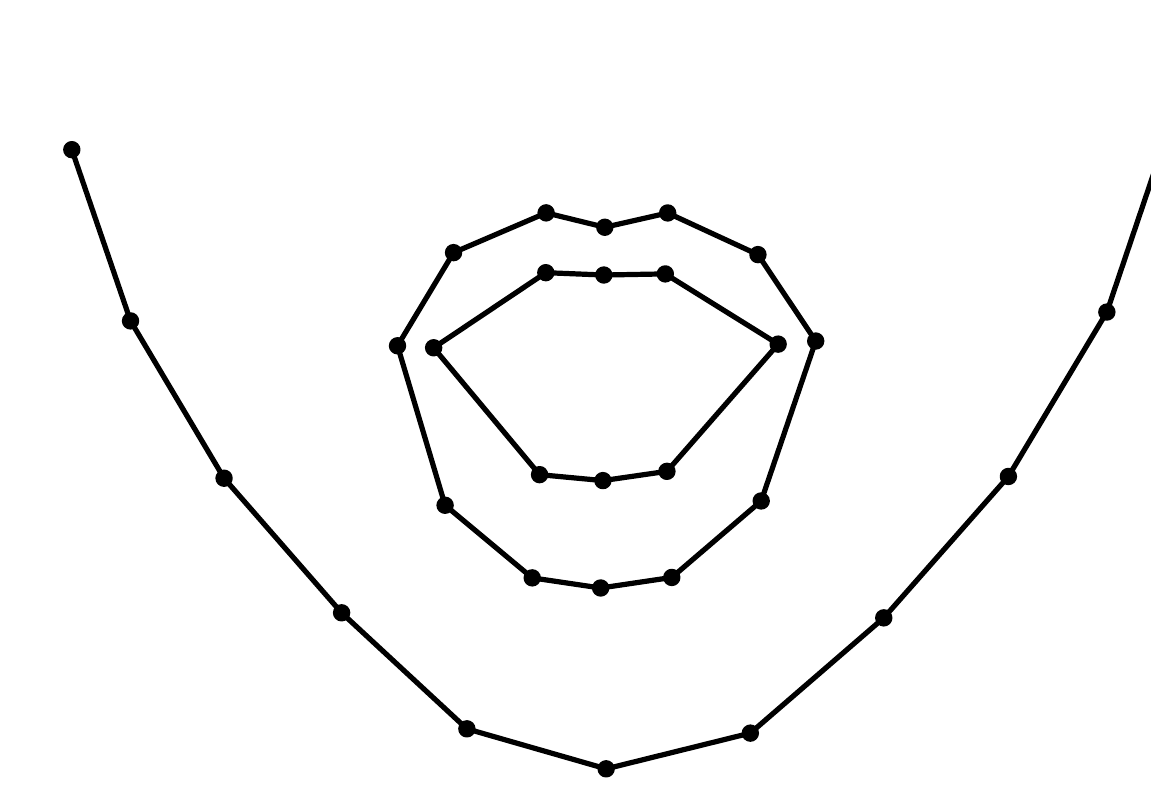}\\
  scale $= 0.1$ & scale $= 0.5$ & scale $= 0.9$ & scale $= 1.1$ & scale $= 1.5$ & scale $= 1.9$
  \end{tabular}
  \caption{Illustration of the effect of scaling the PCA values of the facial landmarks for the same video frame.}
  \label{fig:scaling_factors}
\end{figure*}


\section{Dataset and Data Preparation}\label{sec:dataset}
We use the GRID corpus~\cite{cooke2006audio}, which contains multimodal recordings of $34$ native English speakers reciting fixed grammar sentences of the form, \emph{command-color-preposition-letter-number-adverb}, where each token in the grammar has a limited number of options (see~\cite{cooke2006audio} for details).  An example utterance might be ``place blue at A 3 now''.  The dataset contains all combinations of the colors, letters, and numbers.  Each speaker recited a total of $1000$ sentences, and each video is three seconds long recorded at $25$ frames per second.

We use Dlib~\cite{king2009dlib} to extract $68$ facial landmarks from each video frame and then reduce the effects of tracking noise by convolving the sequence across time with an averaging window of width three frames. We then align the facial landmarks to a common reference~\cite{eskimez2018generating} to remove translation, rotation, and scaling variation.

To manipulate the degree of speech articulation, we project the landmarks onto a speaker-specific principal components analysis (PCA) model, and then for each utterance we randomly sample four scaling factors from \{$0.1$, $0.3$, $0.5$, $0.7$, $0.9$, $1.0$, $1.1$, $1.3$, $1.5$, $1.7$, $1.9$\}.  These scaling factors were chosen to have paired values for under- and over-articulation. For example, the values of $0.9$ and $1.1$ are equivalently scaled either side of no change in articulation ($1.0$).  The scaled PCA values were finally projected back to landmarks for use in generating the stimuli for the user-studies.

\section{Subjective Assessment of Visual Speech Quality}\label{sec_assessment}

\textbf{User study setup.} We use the perturbed landmarks from Section~\ref{sec:dataset} to generate the stimuli for three experiments which assess the perceptual quality of visual speech.  In these experiments, viewers were shown pairs of videos: a reference video (no scaling) and a (possibly) modified version of the same sequence containing under- or over-articulated lip motion.  The order of the videos within a pair was always the same (reference video on the left and sample video on the right).  Annotators were asked to indicate using a seven-point
Likert scale if the sample animation looked: \{Extremely Worse, Moderately Worse, Slightly Worse, The Same, Slightly
Better, Moderately Better, Extremely Better\} compared to the reference animation.  All annotators spoke English, based on the crowd-sourcing platform's screening process.

\textbf{Data analysis.} We analyzed annotator perceptual judgements in all user studies using linear mixed-effects regression.\footnote{We follow the recommendation by~\cite{saurolewis2021userstats} on using linear regression for ordinal scale data.} For the analysis, we decompose the scaling factor into two categorical variables: articulation type, representing the direction of scaling, and the scaling step. For example, a scaling step of $1$ corresponds to scaling the PCA values by $1.1$ for over-articulation and scaling the PCA values by $0.9$ for under-articulation; a scaling step of $2$ corresponds to scaling the PCA values by $1.3$ for over-articulation and scaling the PCA values by $0.7$ for under-articulation; and so on. Trials with a scaling factor of $1.0$ (ground truth) were removed from this analysis since the scaling step did not apply. Scaling step was sliding difference coded.\footnote{Sliding difference coding compares the mean of the dependent variable for one level of the categorical variable to the mean of the dependent variable for the preceding adjacent level (e.g., scaling step $2$ vs. scaling step $1$).} Articulation type was sum-coded (under-articulation $=1$). All models include the maximal converging random effects structure for the experimental design (a random intercept for annotator and a random intercept for item nested within a ground truth sequence, relating to the version of the ground truth sequence for Experiment 1; random intercepts for annotator, item, and speaker for Experiments 2 and 3).  

\subsection{Experiment 1: Rendering on Point-light Displays}\label{sec_landmarks}

Our first user study aims to measure the impact of under-articulated and over-articulated lip motion on the perceived quality of visual speech rendered as landmarks.

To remove inter-speaker variability effects from this initial analysis, we select $996$ sequences for speaker \texttt{s-25}.\footnote{The landmark extraction failed for videos: \texttt{prat5s}, \texttt{pwin2n}, \texttt{bbwq4n}, and \texttt{brwk8p}.} This speaker was selected as they have the largest variance in lip-opening height, measured as the distance between the mid-points of the upper and lower lips.  To ensure we focus exclusively on speech-related facial motion when assessing the quality of visual speech, we remove landmarks that correspond to the eyes, nose, and eyebrows. As a result, each frame is represented by $31$ landmarks, where each landmark is a pair of \textit{x}-\textit{y} points, which were rendered to videos.  See Figure~\ref{fig:scaling_factors} for example rendered video frames.  The videos were presented to annotators as described in the setup.

\begin{table}[t]
\caption{Model summary for the analysis of perceptual judgements for landmarks-only representation. See Section~\ref{subsec:results_study1} for model details. Step is scaling step ($1$-$5$, sliding difference coded); Art. is articulation type (sum-coded, under-articulation $=1$). }
  \label{tab:regression_results}
  \centering
  \resizebox{\columnwidth}{!}{
  \begin{tabular}{lccc}
    \toprule
    \textbf{} & \textbf{Estimate} &   $\mathbf{t}$ &  $\mathbf{p}$\\
    \midrule
    Art.        & $-0.472$  & $-53.53$ & $<0.0001$ \\
    Step $2$ vs. $1$       & $-0.09$  & $-3.07$ & $0.002$\\
    Step $3$ vs. $2$       & $-0.27$  & $-9.75$ & $<0.0001$\\
    Step $4$ vs. $3$       & $-0.56$  & $-20.19$ & $<0.0001$\\
    Step $5$ vs. $4$       & $-0.58$  & $-21.21$ & $<0.0001$\\
    Step $2$ vs. $1$ * Art.       & $-0.16$ & $-4.43$ & $<0.0001$\\
    Step $3$ vs. $2$ * Art.         & $-0.18$ & $-6.39$ & $<0.0001$\\
    Step $4$ vs. $3$ * Art.        & $-0.32$ & $-11.75$ & $<0.0001$\\
    Step $5$ vs. $4$ * Art.         & $-0.43$ & $-15.76$ & $<0.0001$\\
    \bottomrule
  \end{tabular}
  }
\end{table}

\subsubsection*{Results}\label{subsec:results_study1}



\begin{figure}[t]
  \centering
  \includegraphics[width=\columnwidth]{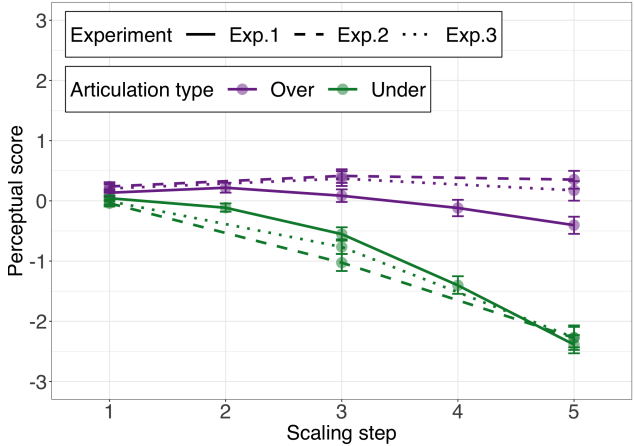}
  \caption{Mean perceptual scores for over- (purple) and under-articulated (green) visual speech under varying scaling steps in the three experiments. A perceptual score of $-3$, $-2$, $-1$, $0$, $1$, $2$, and $3$ on the $y$-axis corresponds to ``Extremely Worse'', ``Moderately Worse'', ``Slightly Worse'', ``The Same'', ``Slightly Better'', ``Moderately Better'', and ``Extremely Better'', respectively. The solid lines represent scores for point-light displays (Exp.1) (Section~\ref{subsec:results_study1}), the dashed lines represent scores for the photo-realistic model driven by ground-truth landmarks (Exp.2) (Section~\ref{sec_photo_realistic_gt}), and the dotted lines represent the photo-realistic model driven by landmarks predicted from speech (Exp.3) (Section~\ref{sec_photo_realistic_pred}). The error bars represent bootstrapped $95$\% confidence intervals.}
  \label{fig:crowd_sourcing_results}
\end{figure}

Across all scaling steps, annotators preferred over-articulated lip motion (see Table~\ref{tab:regression_results}). Every scaling step increase resulted in a lower perceptual score compared to the previous step, suggesting that the more the degree of speech articulation is modified, the more negative impact it has on the perception of its quality. However, articulation type interacted with scaling step---an increase in scaling had a more pronounced effect on under-articulation compared to over-articulation. Interestingly, over-articulation was rated as high as the ground truth, or even better, at all but the highest scaling steps (see Figure ~\ref{fig:crowd_sourcing_results}).
\begin{table}[t]
\caption{Model summary for the analysis of perceptual judgements for photo-realistic talking face. See Section~\ref{study_2_results} for model details. Step is scaling step (sliding difference coded, $3$ vs. $1$; $5$ vs. $3$); Art. is articulation type (sum-coded, under-articulation $=1$); Exp. is Experiment (sum-coded, Exp.2 $=1$); speaker gender is sum-coded, male $=1$.}
  \label{tab:regression_results2}
  \centering
  \resizebox{\columnwidth}{!}{
  \begin{tabular}{lccc}
    \toprule
    \textbf{} & \textbf{Estimate} &  $\mathbf{t}$ &  $\mathbf{p}$\\
    \midrule
    Art.       & $-0.73$ & $-16.41$ & $<0.0001$ \\
    Step $3$ vs. $1$       & $-0.35$ & $-15.69$ & $<0.0001$\\
    Step $5$ vs. $3$       & $-0.75$  & $-32.94$ & $<0.0001$\\
    Exp.       & $0.01$ & $0.44$ & $0.66$\\
    Speaker gender       & $0.01$ & $1.35$ & $0.18$\\
    Step $3$ vs. $1$ * Art.      & $-0.59$ & $-26.32$ & $<0.0001$\\
    Step $5$ vs. $5$ * Art.        & $-0.66$  & $-28.83$ & $<0.0001$\\
    Exp. * Art.       & $0.05$ & $6.67$ & $<0.000$\\
    Step $3$ vs. $1$ * Exp.        & $0.02$ & $1.33$ & $0.18$\\
    Step $5$ vs. $3$ * Exp.       & $-0.07$ & $-4.06$ & $<0.0001$\\
    Step $3$ vs. $1$ * Art. * Exp.     & $0.06$ & $3.35$ & $<0.0001$\\
    Step $5$ vs. $3$ * Art. * Exp.     & $-0.01$ & $-0.83$ & $0.40$\\
    
    \bottomrule
  \end{tabular}
   }
\end{table}

These findings suggest that while viewers are sensitive to the degree of visual speech articulation, the sensitivity has less of an effect for over-articulation compared to under-articulation, and that viewers generally prefer over-articulation.

\subsection{Rendering as Photo-realistic Sequences}\label{sec_photo_realistic}
We use the framework proposed by Zhou et al.~\cite{zhou2020makelttalk} to generate photo-realistic sequences corresponding to dampened/exaggerated visual speech.
This framework generates a video sequence corresponding to a talking face in two stages:  first facial landmarks are predicted from speech,  then the landmarks are used to animate a reference still-image using an image2image translation module similar to the model proposed by Zakharov et al.~\cite{zakharov2019few}. This two-stage pipeline allows first testing photo-realistic sequences generated from scaled ground-truth speech articulation and from scaled landmark sequences that are predicted from the entire talking face pipeline.

\subsubsection{Experiment 2: Using ground-truth landmarks}\label{sec_photo_realistic_gt}

We first randomly sample $30$ utterances from each speaker in the GRID corpus to obtain $990$ utterances.\footnote{Video data for \texttt{s-21} is missing from the corpus.} For the selected utterances, we scale the degree of articulation in the visual speech, as described in Section~\ref{sec:dataset}, but using only scaling factors: \{$0.1$,  $0.5$,  $0.9$, $1.0$, $1.1$,  $1.5$,  $1.9$\} since a power analysis of the results in Section~\ref{subsec:results_study1} showed we would still maintain $100$\% power with this reduced set. The stimuli for the user-study are then created by reconstructing the landmarks from the scaled PCA values and inputting these landmarks with the first video frame from the original sequence to the image2image translation module.

\subsubsection{Experiment 3: Using landmarks predicted from speech}\label{sec_photo_realistic_pred}

For the same utterances in Section~\ref{sec_photo_realistic_gt}, we compute the predicted landmarks from acoustic speech using the pre-trained model provided by Zhou et al.~\cite{zhou2020makelttalk}.  These are then subject to the same scaling as in Section~\ref{sec:dataset}.  The difference between this experiment and that in Section~\ref{sec_photo_realistic_gt} is that the landmarks here are subject to all inaccuracies present in a state-of-the-art talking face system, and we can determine whether the under- vs.\ over-articulation findings still hold.  In other words, does the conclusion that people prefer over-articulation still hold if the over-articulation emphasizes errors made by the model that predicts landmark sequences?

\subsubsection*{Results}\label{study_2_results}

Overall, the results from the experiments with photo-realistic  sequences have similar trends to the results from the experiment that used point-light displays (see Table~\ref{tab:regression_results2} and Figure~\ref{fig:crowd_sourcing_results}). Applying an increasingly larger magnitude scaling to increase the degree of under- or over-articulation resulted in lower perceived quality. Over-articulation was preferred overall, and the preference becomes more pronounced at each increased scaling step. Overall the perceptual scores for the photo-realistic sequences did not differ when either ground-truth landmarks or predicted landmarks were used. However, the difference between under- and over-articulation was more pronounced for the ground-truth landmarks, but this difference was only present at higher scaling steps, suggesting that this difference is somewhat small. Speaker gender did not influence perceptual scores.

We run an additional analysis where we compare the perceptual scores for the point-light display videos (using scaling steps $1$, $2$, and $3$) and the photo-realistic sequences (collapsing over the predicted and ground-truth landmarks). Photo-realistic sequences were rated higher overall. Additionally, the difference between under- and over-articulation was larger for photo-realistic sequences than for videos using point-light displays ($p<0.001$ for all).

The findings from these experiments suggest that the preference for over-articulation generalizes from point-light displays to photo-realistic stimuli of different speakers, and that this preference is stronger for photo-realistic stimuli. Additionally, we find that the perception of photo-realistic sequences does not differ based on whether image2image model is conditioned on ground-truth or predicted landmarks.


\section{Conclusions}
In this work, we studied the relationship between the degree of articulation and the perceived quality of visual speech. We showed that varying the strength of over- and under-articulation affects the perceived quality. Specifically, we found that individuals consistently preferred visually over-articulated speech to visually under-articulated speech for all scaling steps. We also found that while the perceptual score goes down for high scaling steps, the reduction in the score is more pronounced for visually under-articulated speech. The observed preference for over-articulation in point-light displays is even more pronounced for more photo-realistic sequences across different speakers. Finally, the perception of photo-realistic sequences does not differ based on whether the image2image network is driven by ground-truth or predicted landmarks.

\section{Future Work}

The findings from this work impact lip animation models in two aspects. First, the findings from our perceptual studies can be incorporated into training algorithms to yield more natural animations. Specifically, loss functions for neural-based lip-sync systems can be re-weighted to account for the perceived effects of articulation errors (i.e., err on the side of over-articulation for the best perceived outcome). Second, our findings shed light on how lip motion can influence subjective evaluation during model development and bench-marking.
Future work will investigate the effect of other common augmentations (e.g., jitter, synchronization) on the perceived quality of lip motion and will also investigate the automatic prediction of the perceptual scores.

\section{Acknowledgements}
Thanks to our colleagues, Vikram Mitra, Russ Webb, and Ahmed Hussen Abdelaziz, for their insightful feedback on this work. Also, thanks to Lukas Michelsen, the annotation team, and the annotators for their help in conducting the perceptual studies.


\bibliographystyle{IEEEbib}
\bibliography{refs}

\end{document}